\documentclass{sf2a-conf2018}
\usepackage{graphicx}
\usepackage{hyperref}
\usepackage[]{natbib}  
\usepackage{epstopdf}

\def\BibTeX{{\rm B\kern-.05em{\sc i\kern-.025em b}\kern-.08em
    T\kern-.1667em\lower.7ex\hbox{E}\kern-.125emX}}
\bibpunct{(}{)}{;}{a}{}{,}  

\usepackage[utf8]{inputenc}
\usepackage[T1]{fontenc}
\usepackage {amsmath, amsfonts, amssymb}
\usepackage {bm}
\usepackage {mathabx} 
\usepackage {multirow}

\begin{document}

\TitreGlobal{SF2A 2017}


\title{Tidal heating in multilayer planets : \\
Application to the TRAPPIST-1 system}

\runningtitle{Tidal heating in multilayer planets}

\author{S. Breton$^{1,}$}\address{AIM, CEA, CNRS, Université Paris-Saclay, Université Paris Diderot, Sorbonne Paris Cité, F-91191 Gif-sur-Yvette, France}
\address{\'Ecole polytechnique, Route de Saclay, 91120 Palaiseau, France}

\author{E. Bolmont$^{1}$}




\author{G. Tobie}\address{Laboratoire de Planétologie et Géodynamique,UMR-CNRS 6112, Université de Nantes, 44322 Nantes cedex 03, France}

\author{S. Mathis$^{1}$}

\setcounter{page}{237}


\maketitle

\begin{abstract}

TRAPPIST-1 \citep{2017Natur.542..456G} is an extremely compact planetary system: seven earth-sized planets orbit at distances lower than 0.07 AU around one of the smallest M-dwarf known in the close neighborhood of the Sun (with a mass of less than $0.09~M_\odot$). 
With 3 planets within the classical habitable zone, this system represents an interesting observational target for future instruments such as the JWST \citep[e.g.][]{2016MNRAS.461L..92B}.

As the planets are close-in, tidal interactions play a crucial role in the evolution of the system by controlling both orbital configurations and rotational states of the planets. 
For the closest planets, the associated tidal dissipation could have an influence on their internal evolution and potentially on their climate and habitability \citet{2018A&A...612A..86T}.

Following \citep{2005Icar..177..534T}, we build multilayer models of the internal structure of the TRAPPIST-1 planets accounting for the mass and radius of \citet{2018A&A...613A..68G}, then we compute the tidal response and estimate the tidal heat flux of each planet as well as the profile of tidal heating with depth. 
Finally, we compare our results to the homogeneous model of \citet{2012CeMDA.112..283E} and assess the impact heating rate on the thermal state of each layer of the planet.

\end{abstract}


\section {Introduction: TRAPPIST-1, system of interest}

The  TRAPPIST-1  planets  represent good  candidates  for  exobiology  studies.  
Thus,  it  is  important  to  constrain  the  system  to prepare for future observations. 
For these close-in planets, their orbital, rotational and interior evolution can be strongly driven by tides. 
Unfortunately, most tidal orbital models use only simple tidal models, considering homogeneous bodies \citep{2012CeMDA.112..283E,2018ApJ...857..142M} or using simplified approaches \citep{1981A&A....99..126H}, that are not satisfying for our purpose. 

Following \citet{2005Icar..177..534T}, we use here a multilayer model to describe planet interiors and we use Andrade rheology to compute the response of those interiors to a tidal excitation. 

\section {A multilayer model for planetary structures: comparison with the homogeneous model}

\begin{figure} [!h]
\includegraphics[width = \textwidth]{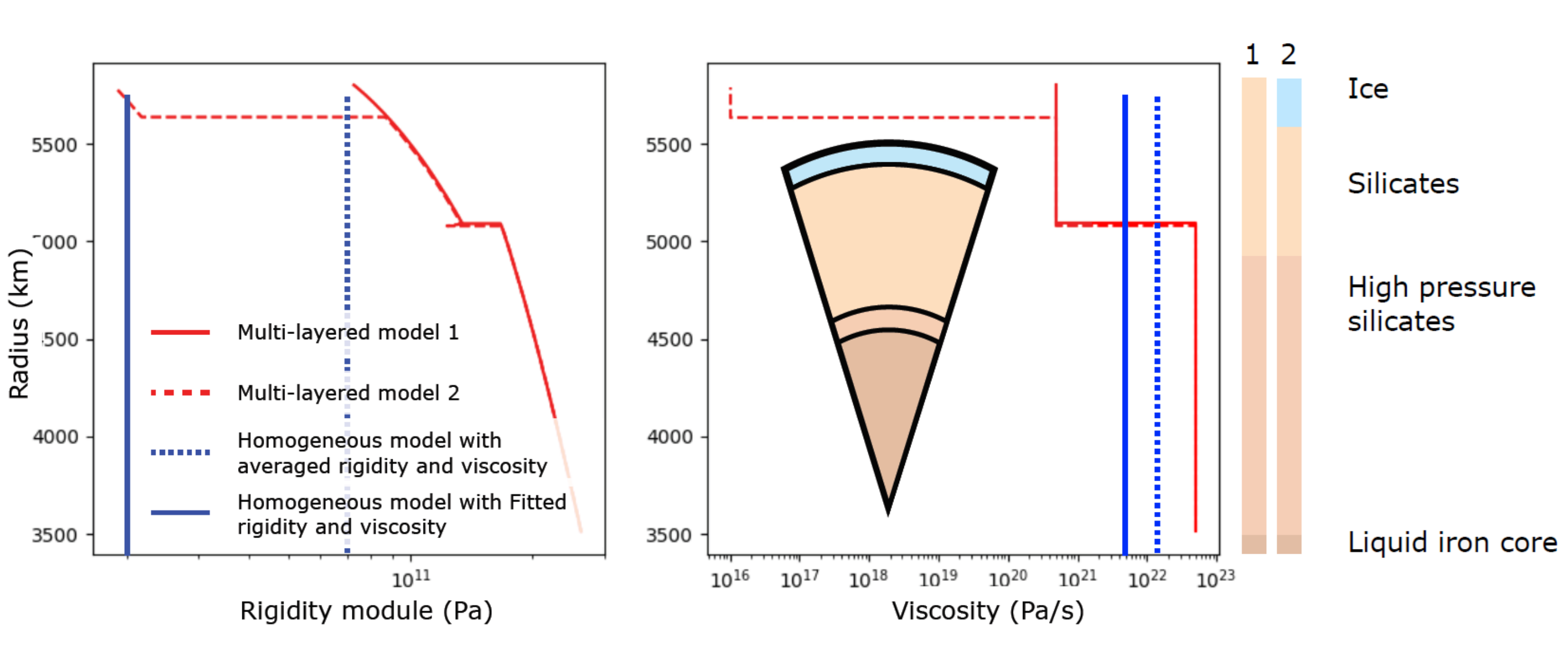}
\caption{Radial profile of the rigidity module and viscosity for TRAPPIST-1e (taken as a planet of 0.772 $M_\Earth$, 0.910 $R_\Earth$, see \citealt{2018A&A...613A..68G}).}
\label{Structure}
\end{figure}

Figure~\ref{Structure} shows the planetary structure used in the multilayer model \citep{2005Icar..177..534T,2007Icar..191..337S}. 
A mantle with radially evolving functions of density and rigidity module encircles a liquid iron core. The physical properties of the mantle are especially influenced by its iron content. 
Additional layers of ice may be present around the mantle. 

The Love number $k_2$ of a body describes its response to a tidal potential. 
In particular, its imaginary part allows to estimate the dissipation due to tides in the planet interior. 
We want to compare the evolution of the imaginary part according to the tidal excitation frequency for different possible planetary models. 
\newpage
To compare multilayer and homogeneous model, we compute 4 internal structure profiles for TRAPPIST-1e:
\begin{itemize}
\item[-] multilayer profile 1 with 138\% ratio Fe/Si in the mantle with respect to Earth and 0\% proportion of ice;
\item[-] multilayer profile 2 with 150\% Fe/Si ratio in the mantle with respect to the Earth and 5\% proportion of ice;
\item[-] homogeneous profile 1 with uniform viscosity and rigidity module, which are mean values of those used for profile 1;
\item[-] homogeneous profile 2 with viscosity and rigidity module values that allow to fit better to the dissipation curve of the multilayer profile 1;
\end{itemize}

\begin{figure} [!t]
\includegraphics[width = \textwidth]{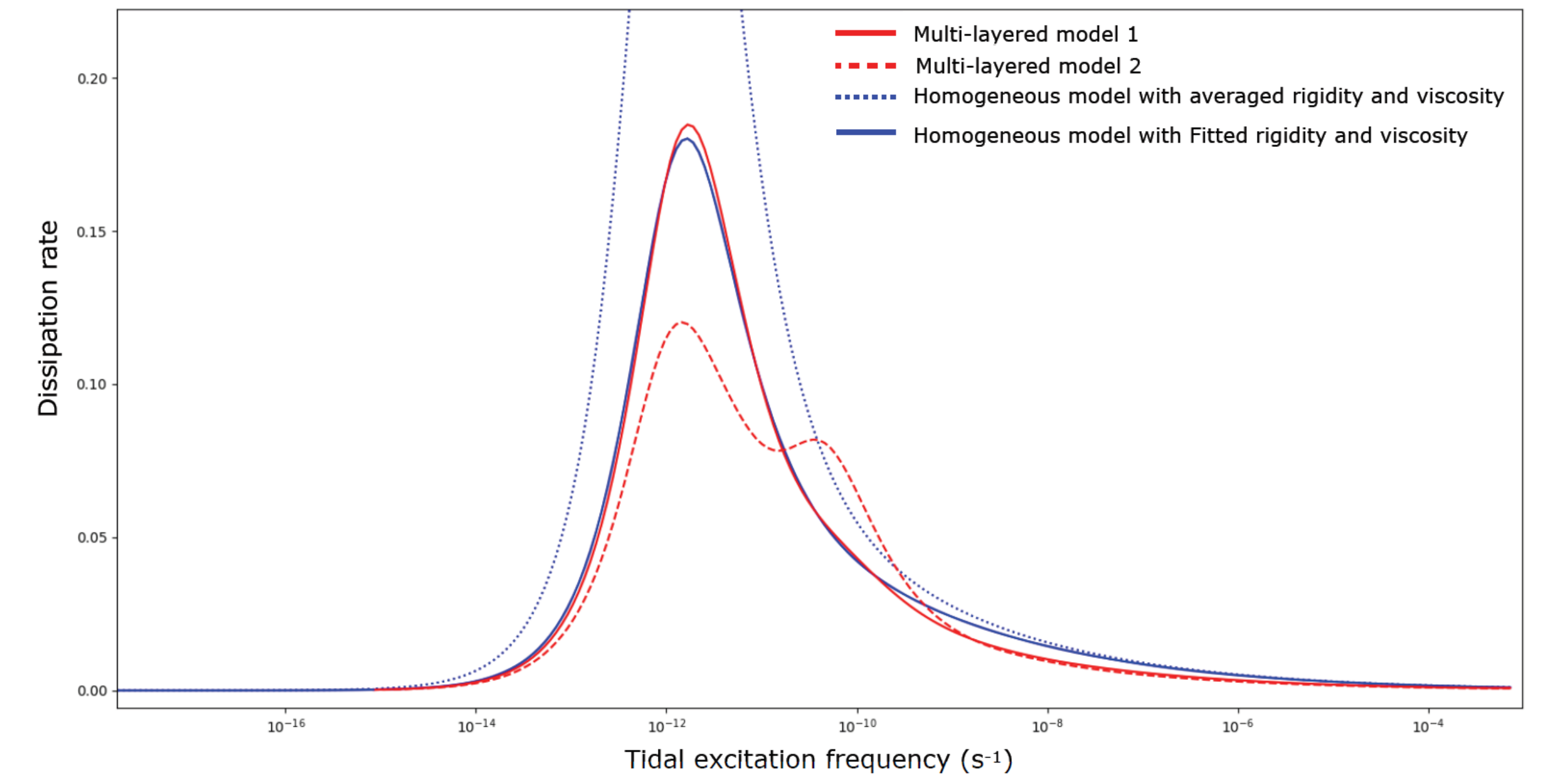}
\caption{Comparison between the value of the dissipation (imaginary part of the Love number) for the  planet  TRAPPIST-1e,  computed  thanks  to  Takeushi  and  Saito’s  multilayer  approach  \citep{TS1972} in red (multi-layer model 1: purely rocky planet, dashed line multi-layer model 2: ice content, see text). The dissipation according to Efroimsky’s formulation for homogeneous spherical bodies \citep{2012CeMDA.112..283E} is in blue.}
\label{Efroimsky} 
\end{figure}

The chosen rheology is Andrade’s one \citep[e.g.][]{2011JGRE..116.9008C}: viscoelastic rheology with memory of the material.
Figure~\ref{Efroimsky} shows that the homogeneous model 1 does not reproduce the behavior of a multilayer planet. It overestimates quite significantly the dissipation.
Figure~\ref{Efroimsky} also shows that the fitted homogeneous model (model 2) does not reproduce the behavior of a planet with ice computed with the multilayer model. We see that with only 5\% of ice the frequency dependency of the imaginary part of the Love number is totally different. In particular, the multilayer model displays a two-peak feature (peak at lower frequencies for rock and at higher frequencies for ice) as well as a difference in amplitude.

Table~\ref{Recap} recapitulates the composition values considered for TRAPPIST-1 planets.

\begin {table} [!h]
\begin{minipage}[c]{.5\linewidth}
\begin {center}
\begin {tabular}{cccc}
\hline \hline
& & Ice (\%) & Fe/Si (\%) \\
\hline
\multirow {2}{*}{b} & b1 & 20.5 & 100 \\
& b2 & 24.0 & 150 \\
\hline
\multirow {2}{*}{c} & c1  & 7.5 & 100 \\
& c2 & 11.0 & 150 \\
\hline
\multirow {2}{*}{d} & d1 & 19.5 & 100 \\
& d2 & 22.5 & 150 \\
\hline
\multirow {2}{*}{e} & e1 & 0.0 & 138 \\
& e2 & 5.0 & 150 \\
\hline
\multirow {2}{*}{f} & f1 & 11.0 & 100 \\
& f2 & 14.5 & 150 \\
\hline
\multirow {2}{*}{g} & g1 & 19.0 & 100 \\
& g2 & 22.5 & 150 \\
\hline
\multirow {2}{*}{h} & h1 & 10.0 & 100 \\
& h2 & 13.5 & 150 \\
\hline
\end {tabular}
\end {center}
\end{minipage}
\begin{minipage}[c]{.5\linewidth}
\begin {center}
\begin{tabular}{cc|c}
\hline \hline
&& Tidal  \\
&& internal \\
&& heating (TW) \\
\hline
\multirow {2}{*}{b} & b1 & 413 \\
& b2 & 519 \\
\hline
\multirow {2}{*}{c} & c1 & 15 \\
&c2 & 18 \\
\hline
\multirow {2}{*}{d} & d1 & 3.6 \\
&d2 & 4.5 \\
\hline
\multirow {2}{*}{e} & e1 & 2.8 \\
&e2 & 2.9 \\
\hline
\multirow {2}{*}{f} & f1 & 1.0 \\
&f2 & 1.1 \\
\hline
\multirow {2}{*}{g} & g1 & 0.09 \\
&g2 & 0.11 \\
\hline
\multirow {2}{*}{h} & h1 & 0.0012 \\
&h2 & 0.0014 \\
\hline
\end{tabular}
\end {center}
\end{minipage}

\caption { \label {Recap} }{ {\bf Left :} Compositions considered for each planet. The ice composition is given as a percentage of the total mass. The Fe/Si proportion in the mantle is given using Earth composition as a reference, a Fe/Si ratio of 100 \% corresponding to the Earth ratio. Two possible compositions are considerer for each planet. {\bf Right :} Internal tidal heating for the TRAPPIST-1 planets.}
\end{table}

\section {Tidal heating in TRAPPIST-1 planetary interiors}

\begin{figure} [!b]
\begin{center}
\includegraphics[width = \textwidth]{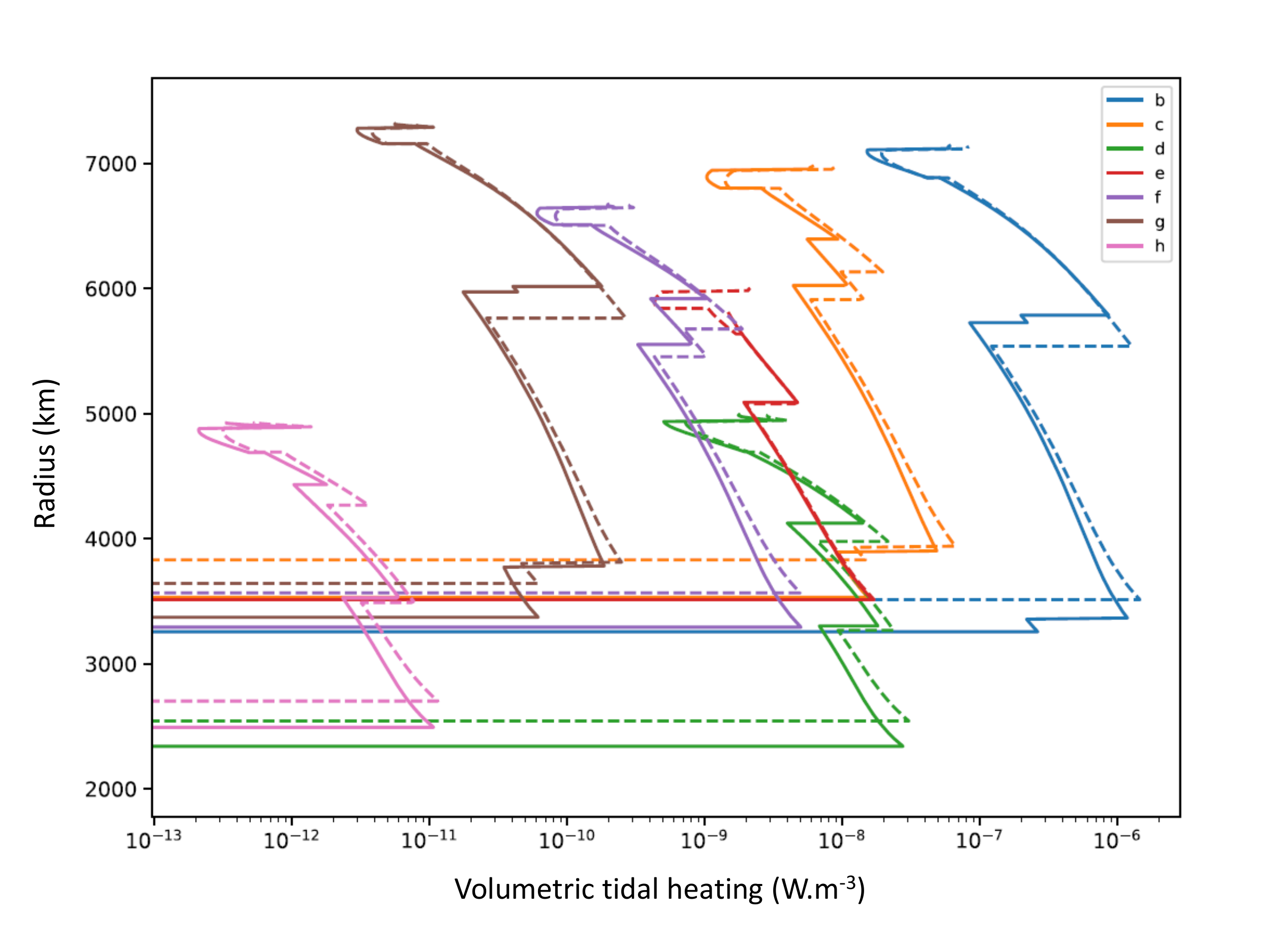}
\caption{Profile of tidal heating for both compositions (see Table~\ref{Recap}) of each planet of TRAPPIST-1.}
\label{TidalHeating}
\end{center}
\end{figure}

The function of volumetric tidal heating $h_{\rm{tide}}$ that we use to compute tidal heating rates profile for TRAPPIST-1 planets is given by the Eq.~14 of \citet{2005Icar..177..534T}. 
This formulation works for coplanar orbits with small eccentricities.
The results are given in Table~\ref{TidalHeating}. Reference global heating values for Io and the Earth are respectively of 100 and 50 TW. 
TRAPPIST-1b is very close to the star and experiences an extremely important heating rate (4 to 5 times more than for Io, the most active telluric body of the Solar system, 8 to 10 times more important than the total heating on Earth).
Even if the heating rate decreases quickly as we go further from the star, for TRAPPIST-1e, one of the planets of the habitable zone, tidal heating could play a role in the energy budget of the atmosphere and influence its habitability. 

\begin{acknowledgements}
E.B. and S.M. acknowledge funding by the European Research Council through ERC grant SPIRE 647383 and support by the CNES PLATO grant at CEA-Saclay. 
The research leading to these results has received funding from the European Union’s Horizon 2020 Research and Innovation Programme, under Grant Agreement 776403 (ExoplANETS A). 
S. Breton was supported by Ecole Polytechnique (Palaiseau, France) and CEA.
\end{acknowledgements}


\bibliographystyle{aa}

\end{document}